# Nuclear Pulse Charge Measurement with a Method of Time over Linear Threshold

Zhengqi Song, Yonggang Wang, Yong Xiao, Qiang Cao

*Abstract*—Time over dynamic threshold (TODT) method, proposed in our previous work, has been successfully used for nuclear pulse charge measurement in PET detectors. It has advantages of strict linearity, large dynamic range, and better energy resolution, but requires a relatively complex circuit to generate specific dynamic threshold. In this paper, we propose to replace the dynamic threshold by a simpler linearly increasing threshold (called as TOLT method) with the aim to simplify the threshold generation circuit meanwhile maintaining its high energy resolution. Mathematical analysis on this replacement and the related realization circuit are presented. By energy spectrum measurement of PET detectors, the method is evaluated. The energy resolutions of PET detectors, composed of a PMT coupled with LYSO and LaBr3 crystal, are measured as 12.54% and 5.18% respectively, which is equivalent to the result obtained by TODT method. The test results show that the TOLT method is more practicable for charge measurement of nuclear detectors.

## I. INTRODUCTION

In TODT measurement, a specific dynamic threshold is used to make the relation between TODT and pulse height subject to strict linearity. Hence measured energy resolutions are better compared to TOT measurement [1]. A high-speed DAC was applied to generate dynamic threshold (DT) in our initial TODT design [2]. But it could add complexity to the system and introduce in more factors affecting performance. So a new idea, using simple analog circuit to generate DT, had been evaluated and tested in [3], it brought us lower power consumption and higher integration. This idea inspired us to replace specific DT with simpler linear increasing threshold, because simpler threshold means simpler analog circuit.

But a linear threshold (LT) could undermine the linearity of the measurement. According to the Mathematical analysis in this paper, a post-correction can offset this non-linearity. Thus the measured energy resolutions won't be compromised. Circuits are built to verify that. Test results show that strict linearity and high energy resolutions can be obtained. So, this method (called as TOLT method) will come in handy for applications like PET that require high integration and low cost.

This manuscript received January 30, 2018. This work was supported by the National Natural Science Foundation of China (NSFC) under Grants 11475168 and 11735013.

The authors are with the Department of Modern Physics, University of Science and Technology of China, Hefei, Anhui, 230026, China. (Corresponding author: wanustc.edu.cn).

## II. TEST CIRCUIT AND MATHEMATICAL INTERPRETATION

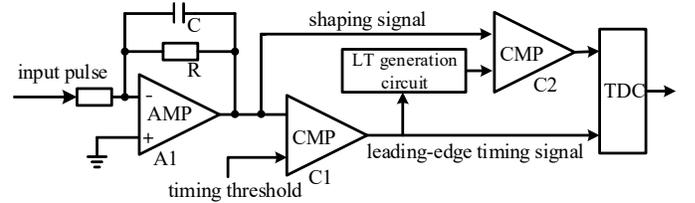

Fig .1. Test circuit for TOLT method

In Fig .1, assume a pulse signal is coming from PMT, after passing through a shaping circuit built by A1 and RC, the shaping signal will be fed into two comparators. One of them (C1) is for leading-edge timing. The output of C1 is lead-edge timing signal that triggers LT to ramp from the ground after a certain delay. Fig .2 shows the circuit for LT generation. The delay is determined by the route in circuit so it's a fixed value.

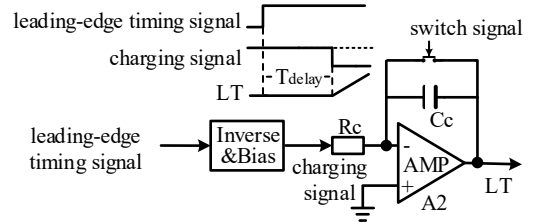

Fig .2. Analog circuit for LT generation

Another comparator, C2, flips when the voltage of LT is over the voltage of shaping signal, we call this moment as TOLT (time over linear threshold). Time stamp of TOLT will be recorded by a high-precision (RMS 40ps) TDC. Fig .3 shows the conception of TOLT measurement.

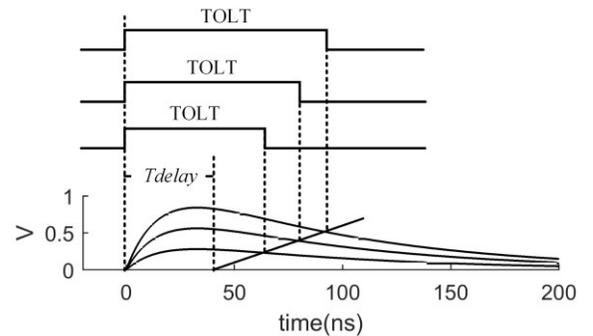

Fig .3. Conception of TOLT measurement.

Shaping signal can be expressed as (1) if we set leading-edge time as zero point. Value of E represents pulse charge.

$$U_{shaping} = E(e^{-t/\tau_2} - e^{-t/\tau_1}) \quad (1)$$

When threshold starts to ramp from the 0, the slope is K:

$$U_{LT} = K(t - T_{delay})$$

At the moment of TOLT, the LT equals to shaping signal:

$$E(e^{-TOLT/\tau_2} - e^{-TOLT/\tau_1}) = K(TOLT - T_{delay})$$

$$E \propto \frac{(TOLT - T_{delay})}{(e^{-TOLT/\tau_2} - e^{-TOLT/\tau_1})}$$

Or it can be written as:

$$\text{pulse charge} \propto \frac{TOLT - T_{delay}}{(e^{-TOLT/\tau_2} - e^{-TOLT/\tau_1})} \quad (2)$$

$T_{delay}, \tau_2, \tau_1$ are all constants during the test, they can be measured advanced. With (2), the relationship between TOLT and pulse charge can be corrected to strict linearity. So the measured energy resolution can reach a high level.

## III. TEST RESULTS

We have built test circuit in Fig .1 and Fig .2. ADA4899 from ANALOG DEVICES has been adopted for A1, ADCMP602 for C1 and C2, and AD8066 for A2. Fig .4 shows the linearity measured from the test circuit. Pulses with different height are generated by an Arbitrary Function Generator (AFG3252) from Tektronix. TDC built in cyclone III FPGA is used to measure TOLT. Measured relation between TOLT and pulse height has been draw out in Fig .4(a). After correction had been performed, relation between corrected TOLT and pulse height yields strict linearity as shown in Fig. 4(b).

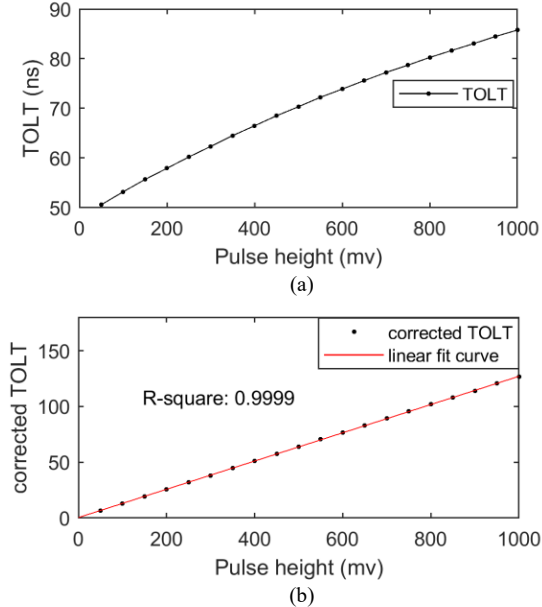

Fig .4.(a) relationship between TOLT and pulse height; (b)relationship between pulse height and corrected TOLT.

Then the energy spectrum of LaBr3 and LYSO had been measured. Na-22 had been used as radioactive source. gamma rays will be detected by PMT R9800 from Hamamatsu, which outputs pulses to our test circuit.

After correction, energy spectrums are shown in Fig .5. Gaussian Fitting had been used for peak at 511Kev. Energy resolutions, represented by FWHM/511Kev, equals to 5.18% and 12.54% for LaBr3 and LYSO respectively. The measured energy resolutions are reasonably fine compared to TODT method.

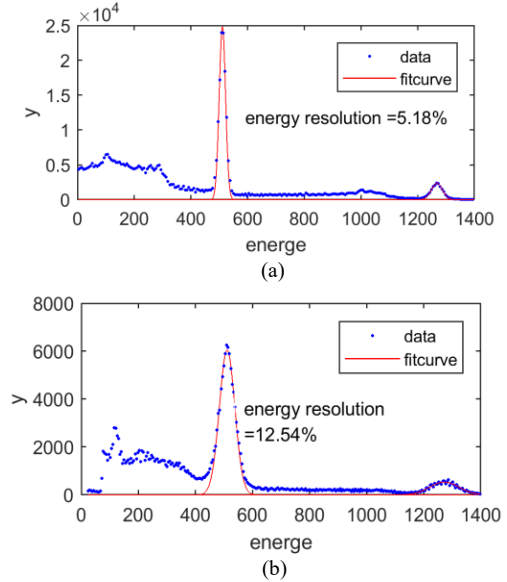

Fig .5. measured energy spectrum of (a):LaBr3, (b):LYSO

## IV. CONCLUSION

By adding a correction table, using a linearly increasing threshold to compare with the shaping signal from nuclear detectors could transfer the nuclear charge into time width linearly. The key point of the TOLT method is to keep the cross position high without losing the measurement dynamic range, which maintains a high signal-to-noise ratio for gaining high energy resolution.

Combining FPGA with the front-end circuit brought some new prospects to the TOLT method. The LT generation circuit in this paper, though simple, can be simplified further by using I/O port on FPGA as constant current source. Other components, like TDC [4] and correction table, can all be easily constructed in FPGA without adding complexity to the system. To fully exploit advantages of FPGA, schemes with high integration and low cost would be easier to realize. This points out the direction to our future work: in front-end circuit, the pursuit of higher integration level will never stop.